%% file: issre23.tex
\documentclass[conference]{IEEEtran}
\IEEEoverridecommandlockouts
\usepackage[table,xcdraw]{xcolor}
\usepackage{booktabs} 
\usepackage{amsmath,amssymb,amsfonts}
\usepackage{algorithmic}
\usepackage{graphicx}
\usepackage{textcomp}
\usepackage{xcolor}
\usepackage{url}
\usepackage{pifont}
\usepackage{enumitem}
\usepackage[caption=false,font=footnotesize,labelfont=sf,textfont=sf]{subfig}
\usepackage{makecell}
\usepackage{flushend}
\usepackage{fancyvrb}
\usepackage{verbatim}
\usepackage[most]{tcolorbox}
\usepackage{balance}
\usepackage{marvosym}
\usepackage{hyperref}
\definecolor[named]{ACMBlue}{cmyk}{1,0.1,0,0.1}
\definecolor[named]{ACMYellow}{cmyk}{0,0.16,1,0}
\definecolor[named]{ACMOrange}{cmyk}{0,0.42,1,0.01}
\definecolor[named]{ACMRed}{cmyk}{0,0.90,0.86,0}
\definecolor[named]{ACMLightBlue}{cmyk}{0.49,0.01,0,0}
\definecolor[named]{ACMGreen}{cmyk}{0.20,0,1,0.19}
\definecolor[named]{ACMPurple}{cmyk}{0.55,1,0,0.15}
\definecolor[named]{ACMDarkBlue}{cmyk}{1,0.58,0,0.21}
\hypersetup{colorlinks=true,
  linkcolor=ACMRed,
  citecolor=ACMPurple,
  urlcolor=ACMDarkBlue,
  filecolor=ACMDarkBlue}

\makeatletter 
\newcommand{\linebreakand}{%
  \end{@IEEEauthorhalign}
  \hfill\mbox{}\par
  \mbox{}\hfill\begin{@IEEEauthorhalign}
}
\makeatother 

\usepackage{amsmath}  

\begin{document}

\title{Loghub: A Large Collection of System Log Datasets for AI-driven Log Analytics}


\author{
\IEEEauthorblockN{Jieming Zhu\IEEEauthorrefmark{1}\thanks{\hspace{-2ex}\IEEEauthorrefmark{1} The work was done when the authors were affiliated with CUHK.}, Shilin He\IEEEauthorrefmark{1}, Pinjia He\IEEEauthorrefmark{2}\textsuperscript{\Letter}\thanks{\hspace{-2ex}\textsuperscript{\Letter} Pinjia He is the corresponding author.}, Jinyang Liu\IEEEauthorrefmark{3}, Michael R. Lyu\IEEEauthorrefmark{3}}\vspace{1ex}
\IEEEauthorblockA{
\IEEEauthorrefmark{2}School of Data Science, The Chinese University of Hong Kong, Shenzhen (CUHK Shenzhen), China\\
\IEEEauthorrefmark{3}Department of Computer Science and Engineering, The Chinese University of Hong Kong, China\\ 
jiemingzhu@ieee.org~~
slhe@link.cuhk.edu.hk~~
hepinjia@cuhk.edu.cn~~
\{jyliu, lyu\}@cse.cuhk.edu.hk
\vspace{-1ex}
}
}



\maketitle

\begin{abstract}
Logs have been widely adopted in software system development and maintenance because of the rich runtime information they record. In recent years, the increase of software size and complexity leads to the rapid growth of the volume of logs. To handle these large volumes of logs efficiently and effectively, a line of research focuses on developing intelligent and automated log analysis techniques. However, only a few of these techniques have reached successful deployments in industry due to the lack of public log datasets and open benchmarking upon them. To fill this significant gap and facilitate more research on AI-driven log analytics, we have collected and released loghub, a large collection of system log datasets.
In particular, loghub provides 19 real-world log datasets collected from a wide range of software systems, including distributed systems, supercomputers, operating systems, mobile systems, server applications, and standalone software. In this paper, we summarize the statistics of these datasets, introduce some practical usage scenarios of the loghub datasets, and present our benchmarking results on loghub to benefit the researchers and practitioners in this field. Up to the time of this paper writing, the loghub datasets have been downloaded for roughly 90,000 times in total by hundreds of organizations from both industry and academia. The loghub datasets are available at \href{https://github.com/logpai/loghub}{https://github.com/logpai/loghub}.

\end{abstract}


\input{Sections/introduction}
\input{Sections/dataset}

\input{Sections/application}

\input{Sections/casestudy_parsing}

\input{Sections/casestudy_compression}
\input{Sections/casestudy_ad}

\input{Sections/relatedwork}
\input{Sections/conclusion}

\balance
\bibliographystyle{IEEEtranS}
\bibliography{issre23}

\end{document}

%% file: Sections/introduction.tex
\section{Introduction}\label{sec:intro}

Logs have been widely adopted in software system development and maintenance~\cite{log_survey, log_empirical_study}. In industry, it is a common practice to record detailed software runtime information into logs, allowing developers and operating engineers to track system behaviors and perform post-mortem analysis.

In general, logs are a form of unstructured texts printed by logging statements (e.g., \textit{logging.info()}, \textit{printf()}, \textit{Console.Writeline()}) in source code. A log message, as illustrated in the following example, records a specific system event with a set of fields: \textbf{timestamp} (the occurrence time of the event, \textit{e.g., 2008-11-09 20:46:55,556}), \textbf{verbosity level} (the severity level of the event, \textit{e.g., INFO}), and \textbf{message content} that describes the event in free text.\\

\begin{footnotesize}
\begin{Verbatim}[frame=single]
2008-11-09 20:46:55 INFO dfs.DataNode$PacketRespond: 
Received block blk_3587508140051953248 of size 6710.
\end{Verbatim}
\end{footnotesize}

The rich information recorded by logs enables developers to conduct a variety of log-based analysis and management tasks, such as anomaly detection~\cite{Xu_sosp_2009, ll-25, slheISSRE16, du2017deeplog}, duplicate issue identification~\cite{ding2014mining, lim2014identifying, rakha2018revisiting}, usage statistics analysis~\cite{ll-31}, and program verification~\cite{beschastnikh2011leveraging, verification}. For example, developers could inspect log messages and analyze whether the system behaves as expected. However, software systems are becoming large in scale and complex in structure. The volume of system logs is growing rapidly as well (e.g., 50 GB/hour~\cite{mi2013toward}), making manual log analysis become labor-intensive and time-consuming. To address this problem, a line of research~\cite{Xu_sosp_2009, ll-25, slheISSRE16, ding2014mining, lim2014identifying, rakha2018revisiting, ll-31, beschastnikh2011leveraging, verification} has targeted at making automated log analysis possible based on  artificial intelligence (AI) techniques. These studies demonstrate that the use of AI techniques can greatly facilitate log analysis tasks by extracting critical information of runtime behaviors. 

\begin{figure*}[h]
\centering{} 
\vspace{3pt}
\includegraphics[scale=0.84]{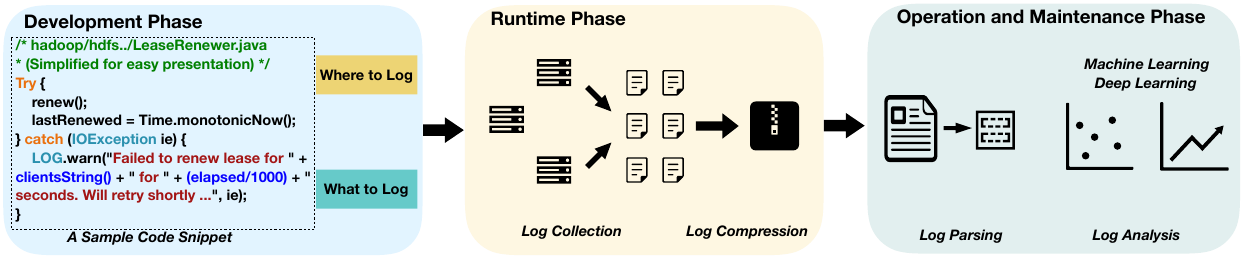}
\caption{Framework of AI-driven Log Analytics}
\label{fig:framework}
\end{figure*}

Figure~\ref{fig:framework} illustrates an overall framework for  AI-driven log analytics. In the development phase, developers can make logging decisions guided by \textit{strategic logging practices} (i.e., "where to log" \cite{qfuICSE14q, zhu15learningToLog} and "what to log" \cite{li17ESEb, he2018characterizing}) mined from high-quality software repositories. At system runtime, logs are \textit{collected and aggregated} in a streaming manner. To reduce the storage cost of system logs, \textit{log compression} techniques~\cite{liu2019logzip} could be further applied. In the operation and maintenance phase, logs need to be parsed into structured events with \textit{log parsing} techniques~\cite{He-DSN,zhu2018tools}, and then facilitate the modeling and mining for a variety of \textit{log analysis} tasks (e.g., anomaly detection~\cite{slheISSRE16}, problem identification~\cite{he2018identifying}). 

Along with this framework, many efforts have been devoted to improving AI techniques towards logging, log collection, log compression, log parsing, and log analysis. Many more methods are being proposed as well. However, there is still a large gap between research and practice. First, researchers in this field often work on their own log data. Logs are scarce data in public for research, because companies are often reluctant to release their production logs due to privacy concerns. Thus, an approach that works well on one type of log data may become ineffective on another type of logs. Second, it is difficult and time-consuming for researchers and practitioners to implement the approaches and accurately reproduce the results without a standard benchmark. 

To bridge this gap, this paper presents loghub, a large collection of system log datasets for AI-driven log analytics. Loghub contains a total of 19 log datasets (see Table~\ref{tab:dataset} for details) generated by a wide range of systems, including distributed systems, super computers, operating systems, mobile systems, server applications, and standalone software. All these logs amount to about 77 GB in total. In particular, some of the logs are production data released from previous studies, while some others are collected from real systems in our lab environment. Among these log datasets, six of them are labeled (e.g., normal or abnormal, alerts or not alerts), which are amendable to studies for anomaly detection and duplicate issues identification. Additionally, other datasets could facilitate research on log parsing, log compression, and unsupervised methods for anomaly detection. Since the first release of these logs, they have been downloaded 90,000+ times by more than 450 organizations from both industry (35\%) and academia (65\%). We envision that loghub could serve as an open benchmarks towards research and practice for AI-driven log analytics. In summary, our work makes the following contributions:
\begin{itemize}
  \item We collect and organize a large collection of log datasets (namely loghub) generated by a wide variety of systems. Loghub consists of 19 datasets, which are valuable for research and practice of AI-driven log analytics (\S~\ref{sec:dataset}).
  \item We introduce practical usage scenarios of the loghub datasets (\S~\ref{sec:application}). We also provide benchmarking results on three typical log analysis tasks using loghub and discuss remaining questions and challenges, shedding light on potential directions for future research and development in log analytics (\S~\ref{sec:benchmarks}).
  \item Our loghub datasets have been made available on Github. Since the release of loghub, they have made a measurable impact to the community, benefiting research in over 450 organizations from both industry and academia.
\end{itemize}


%% file: Sections/dataset.tex
\section{Loghub Datasets}\label{sec:dataset}
\begin{table*}[t]
\renewcommand\arraystretch{1.7}
\setlength{\tabcolsep}{17pt}
\centering{}
\caption{Summary of Loghub Datasets}
\begin{tabular}{|llcrlc|}
\hline
\multicolumn{1}{|c|}{\textbf{Dataset}} & \multicolumn{1}{c|}{\textbf{Description}} & \multicolumn{1}{c|}{\textbf{Time Span}} & \multicolumn{1}{c|}{\textbf{\#Lines}} & \multicolumn{1}{c|}{\textbf{Data Size}} & \textbf{Labeled} \\ \hline
\multicolumn{6}{|c|}{\cellcolor[HTML]{f8f8f8}{Distributed systems}} \\ \hline
\multicolumn{1}{|l|}{HDFS\_v1} & \multicolumn{1}{l|}{Hadoop distributed file system log} & \multicolumn{1}{c|}{38.7 hours} & \multicolumn{1}{r|}{11,175,629} & \multicolumn{1}{r|}{1.47GB} & $\surd$ \\ \hline
\multicolumn{1}{|l|}{HDFS\_v2} & \multicolumn{1}{l|}{Hadoop distributed file system log} & \multicolumn{1}{c|}{N.A.} & \multicolumn{1}{r|}{71,118,073} & \multicolumn{1}{r|}{16.06GB} &  \\ \hline
\multicolumn{1}{|l|}{HDFS\_v3} & \multicolumn{1}{l|}{Instrumented HDFS trace log} & \multicolumn{1}{c|}{N.A.} & \multicolumn{1}{r|}{14,778,079} & \multicolumn{1}{r|}{2.96GB} & $\surd$ \\ \hline
\multicolumn{1}{|l|}{Hadoop} & \multicolumn{1}{l|}{Hadoop mapreduce job log} & \multicolumn{1}{c|}{N.A.} & \multicolumn{1}{r|}{394,308} & \multicolumn{1}{r|}{48.61MB} & $\surd$ \\ \hline
\multicolumn{1}{|l|}{Spark} & \multicolumn{1}{l|}{Spark job log} & \multicolumn{1}{c|}{N.A.} & \multicolumn{1}{r|}{33,236,604} & \multicolumn{1}{r|}{2.75GB} &  \\ \hline
\multicolumn{1}{|l|}{Zookeeper} & \multicolumn{1}{l|}{ZooKeeper service log} & \multicolumn{1}{c|}{26.7 days} & \multicolumn{1}{r|}{74,380} & \multicolumn{1}{r|}{9.95MB} &  \\ \hline
\multicolumn{1}{|l|}{OpenStack} & \multicolumn{1}{l|}{OpenStack infrastructure log} & \multicolumn{1}{c|}{N.A.} & \multicolumn{1}{r|}{207,820} & \multicolumn{1}{r|}{58.61MB} & $\surd$ \\ \hline
\multicolumn{6}{|c|}{\cellcolor[HTML]{f8f8f8}{Super computers}} \\ \hline
\multicolumn{1}{|l|}{BGL} & \multicolumn{1}{l|}{Blue Gene/L supercomputer log} & \multicolumn{1}{c|}{214.7 days} & \multicolumn{1}{r|}{4,747,963} & \multicolumn{1}{l|}{708.76MB} & $\surd$ \\ \hline
\multicolumn{1}{|l|}{HPC} & \multicolumn{1}{l|}{High performance cluster log} & \multicolumn{1}{c|}{N.A.} & \multicolumn{1}{r|}{433,489} & \multicolumn{1}{r|}{32.00MB} &  \\ \hline
\multicolumn{1}{|l|}{Thunderbird} & \multicolumn{1}{l|}{Thunderbird supercomputer log} & \multicolumn{1}{c|}{244 days} & \multicolumn{1}{r|}{211,212,192} & \multicolumn{1}{r|}{29.60GB} & $\surd$ \\ \hline
\multicolumn{6}{|c|}{\cellcolor[HTML]{f8f8f8}{Operating systems}} \\ \hline
\multicolumn{1}{|l|}{Windows} & \multicolumn{1}{l|}{Windows event log} & \multicolumn{1}{c|}{226.7 days} & \multicolumn{1}{r|}{114,608,388} & \multicolumn{1}{r|}{26.09GB} &  \\ \hline
\multicolumn{1}{|l|}{Linux} & \multicolumn{1}{l|}{Linux system log} & \multicolumn{1}{c|}{263.9 days} & \multicolumn{1}{r|}{25,567} & \multicolumn{1}{r|}{2.25MB} &  \\ \hline
\multicolumn{1}{|l|}{Mac} & \multicolumn{1}{l|}{Mac OS log} & \multicolumn{1}{c|}{7.0 days} & \multicolumn{1}{r|}{117,283} & \multicolumn{1}{r|}{16.09MB} &  \\ \hline
\multicolumn{6}{|c|}{\cellcolor[HTML]{f8f8f8}{Mobile systems}} \\ \hline
\multicolumn{1}{|l|}{Android\_v1} & \multicolumn{1}{l|}{Android framework log} & \multicolumn{1}{c|}{N.A.} & \multicolumn{1}{r|}{1,555,005} & \multicolumn{1}{l|}{183.37MB} &  \\ \hline
\multicolumn{1}{|l|}{Android\_v2} & \multicolumn{1}{l|}{Android framework log} & \multicolumn{1}{c|}{N.A.} & \multicolumn{1}{r|}{30,348,042} & \multicolumn{1}{r|}{3.38GB} &  \\ \hline
\multicolumn{1}{|l|}{HealthApp} & \multicolumn{1}{l|}{Health app log} & \multicolumn{1}{c|}{10.5 days} & \multicolumn{1}{r|}{253,395} & \multicolumn{1}{r|}{22.44MB} &  \\ \hline
\multicolumn{6}{|c|}{\cellcolor[HTML]{f8f8f8}{Server applications}} \\ \hline
\multicolumn{1}{|l|}{Apache} & \multicolumn{1}{l|}{Apache web server error log} & \multicolumn{1}{c|}{263.9 days} & \multicolumn{1}{r|}{56,481} & \multicolumn{1}{r|}{4.90MB} &  \\ \hline
\multicolumn{1}{|l|}{OpenSSH} & \multicolumn{1}{l|}{OpenSSH server log} & \multicolumn{1}{c|}{28.4 days} & \multicolumn{1}{r|}{655,146} & \multicolumn{1}{r|}{70.02MB} &  \\ \hline
\multicolumn{6}{|c|}{\cellcolor[HTML]{f8f8f8}{Standalone software}} \\ \hline
\multicolumn{1}{|l|}{Proxifier} & \multicolumn{1}{l|}{Proxifier software log} & \multicolumn{1}{c|}{N.A.} & \multicolumn{1}{r|}{21,329} & \multicolumn{1}{r|}{2.42MB} &  \\ \hline
\end{tabular}
\label{tab:dataset}
\end{table*}

Loghub maintains a collection of system logs, which are freely accessible for research. Some of the logs are production data released from previous studies, while some others are collected from real systems in our lab environment. Wherever possible, the logs are not sanitized, anonymized or modified in any way. All these logs amount to over 77 GB in total. 

Table~\ref{tab:dataset} presents an overview of the loghub datasets with some details about the description, time span, \#Lines, data size, and the label information. Specifically, time span indicates the time range that the logs are collected. \#Lines denotes the total number of log lines in a dataset. Data size shows the uncompressed log volume size. The ``labeled" column indicates whether a dataset is labeled with anomaly information. 

There are two categories of log datasets: \textit{labeled} and \textit{unlabeled}. Logs in labeled datasets contain labels for specific log analysis tasks (e.g., anomaly detection and duplicate issues identification). For example, in the labeled HDFS datasets, the labels indicate whether the system operations on an HDFS block is abnormal. Thus, developers could utilize the labeled HDFS dataset to evaulate their anomaly detection approaches. In loghub, 6 log datasets are labeled, while 13 log datasets are unlabeled. Note that unlabeled log datasets are also useful for the evaluation of log analytics tasks, such as log parsing, log compression, and unsupervised methods (e.g., word2vec). The details of each log dataset in loghub are introduced as follows.

\subsection{Distributed Systems}

\textbf{HDFS.} HDFS is the Hadoop Distributed File System designed to run on commodity hardware. Due to the popularity of HDFS, it has been widely studied in recent years. We provide three sets of HDFS logs in loghub: HDFS-v1, HDFS-v2, and HDFS-v3. {HDFS-v1} is generated in a 203-nodes HDFS using benchmark workloads, and manually labeled through handcrafted rules to identify the anomalies. The logs collected from the work~\cite{ll-42} are sliced into traces (i.e., log sequences) according to block IDs. Then each trace associated with a specific block ID is assigned a ground truth label: \textit{normal} or \textit{abnormal}. Additionally, HDFS-v1 also provide the specific anomaly type information, while futher allows research on duplicate issues identification. {HDFS-v2} is collected by aggregating logs from the HDFS cluster in our lab environment, which comprises one name node and 32 data nodes. The logs are aggregated at the node level. The logs have a huge size (over 16 GB) and are provided as-is without further modification or labeling. {HDFS-v3} is an open dataset from trace-oriented monitoring~\cite{TraceBench}, which is collected through instrumenting the HDFS system using MTracer~\cite{MTracer} in a real IaaS environment. The logs are collected under different workloads (e.g., multiple scales of clusters, different kinds of user requests, various workload levels). In addition to some normal trace logs, abnormal logs are also collected via fault injection. 

\textbf{Hadoop}. Hadoop is a big data processing framework that allows for the distributed processing of large data sets across clusters of computers using simple programming models. Due to the increasing importance of Hadoop in industry, it has been widely studied in the literature. The logs are generated from a Hadoop cluster with 46 cores across five machines in~\cite{lin2016log}. Each machine has Intel(R) Core(TM) i7-3770 CPU and 16 GB RAM. Two testing applications are executed: \textit{WordCount} and \textit{PageRank}. Firstly, the applications are run without injecting any failure. Then, in order to simulate service failures in the production environment, the following deployment failures are injected: (1) \textit{machine down}: during application runtime, turn off one server to simulate the machine failure; (2) \textit{network disconnection}: disconnect one server from the network to simulate the network connection failure; and (3) \textit{disk full}: during application runtime, manually fill up one server's hard disk to simulate the disk full failure. The labels of different failures are provided, making the data amenable to duplicate issues identification research.

\textbf{Spark}. Apache Spark is a unified analytics engine for big data processing, with built-in modules for streaming, SQL, machine learning and graph processing. Currently, Spark has been widely deployed in industry. This dataset was collected by aggregating logs from the running Spark system in our lab environment, which comprises a total of 32 machines. The logs are aggregated at the machine level. The logs have a huge size (over 2 GB) and are provided as-is without further modification or labelling, which involve both normal and abnormal application runs.

\textbf{Zookeeper}. ZooKeeper is a centralized service for maintaining configuration information, naming, providing distributed synchronization, and providing group services. The log dataset was collected by aggregating logs from the ZooKeeper service in our lab environment, which comprises a total of 32 machines, covering a time period of 26.7 days.

\textbf{OpenStack}. OpenStack is a cloud operating system that controls large pools of compute, storage, and networking resources throughout a datacenter. This dataset was provided by~\cite{du2017deeplog} and was generated on CloudLab~\cite{cloudlab}, a flexible, scientific infrastructure for research on cloud computing. Both normal logs and abnormal cases with failure injection are provided, making the data amenable to anomaly detection research. 

\subsection{Supercomputers}

\textbf{BGL}. BGL is an open dataset of logs collected by~\cite{BGLdata} from a BlueGene/L supercomputer system at Lawrence Livermore National Labs (LLNL) in Livermore, California, with 131,072 processors and 32,768 GB memory~\cite{liang2005filtering}. The logs contain alert and non-alert messages identified by alert category tags. In the first column of the log, "-" indicates non-alert messages while others are alert messages. The label information is amenable to alert detection and prediction research.

\textbf{HPC}. HPC is an open dataset from the work~\cite{IPLoM}, containing logs collected from System 20 of the high performance computing cluster at the Los Alamos National Laboratories, which has 49 nodes with 6,152 cores and 128 GB memory per node.

\textbf{Thunderbird.} Thunderbird is an open dataset of logs provided by~\cite{BGLdata}, which was collected from a Thunderbird supercomputer system at Sandia National Labs (SNL) in Albuquerque, with 9,024 processors and 27,072 GB memory. The logs contain alert and non-alert messages identified by alert category tags. In the first column of the log, "-" indicates non-alert messages while others are alert messages. The label information is amenable to alert detection and prediction research.

\subsection{Operating Systems}

\textbf{Windows}. This log dataset was collected by aggregating a number of logs from a lab computer running Windows 7. The original logs were located at \textit{C:/Windows/Logs/CBS}. CBS (Component Based Servicing) is a componentization architecture in Windows, which works at the package/update level. The CBS architecture is far more robust and secure than the installers in previous operating systems. Users benefit from a more complete and controlled installation process that allows updates, drivers and optional components to be added while simultaneously mitigating against instability issues caused by improper or partial installation. The logs have a huge size (over 27 GB) and span a period of 226.7 days.

\textbf{Linux}. Linux logs are usually located at \textit{/var/log/}. The dataset was collected from \textit{/var/log/messages} on a Linux server over a period of 263.9 days, as part of the Public Security Log Sharing Site project~\cite{securitylogproject}.

\textbf{Mac}. We collected the MacOS logs from \textit{/var/log/system.log} on a personal Macbook after 7 days of use. The log records the user activities on the Mac OS. 

\subsection{Mobile Applications}

\textbf{Android}. Android is a popular open-source mobile operating system and has been used by many smart devices. However, Android logs are rarely available in public for research purposes. We provide some Android log files, which were collected on Android smartphones with heavily instrumented modules installed. The Android architecture comprises of five levels, including the Linux Kernel, Libraries, Application Framework, Android Runtime, and System Applications. We provide two dataset versions of Andriod logs printed by the Application Framework: Andriod-v1 and Andriod-v2. The Andriod-v1 dataset is a sampled small log file from Andriod-v2, while in Andriod-v2, the logs cover two types of issues, and each type has over 10 duplicate issue logs. However, due to the high complexity of Android's multi-threading system, it is difficult to pinpoint the abnormal log points.

\textbf{HealthApp}. HealthApp is a mobile application for Andriod devices. We collected the application logs from an Android smartphone after more than 10 days of use.

\subsection{Server Applications}

\textbf{Apache}. Apache HTTP Server is one of the most popular web servers. Apache servers usually generate two types of logs: access logs and error logs. This dataset provides an error log for the purpose of research on anomaly detection and diagnosis. The log file was collected from a Linux system running Apache Web server, as part of the Public Security Log Sharing Site project~\cite{securitylogproject}.

\textbf{OpenSSH}. OpenSSH is the premier connectivity tool for remote login with the SSH protocol. We collected the log from an OpenSSH server in our lab over a period of 28 days.

\subsection{Standalone Software}

\textbf{Proxifier}. Proxifier is a software program, allowing network applications that do not support working through proxy servers to operate through a SOCKS or HTTPS proxy and chains. We collected the Proxifier logs from a desktop computer in our lab.

%% file: Sections/application.tex
\section{Usage of Loghub Datasets}\label{sec:application}
In this section, we present some common usage scenarios of the loghub datasets. 

\subsection{Overview}
\begin{figure}[h]
\centering{} 
\includegraphics[scale=0.9]{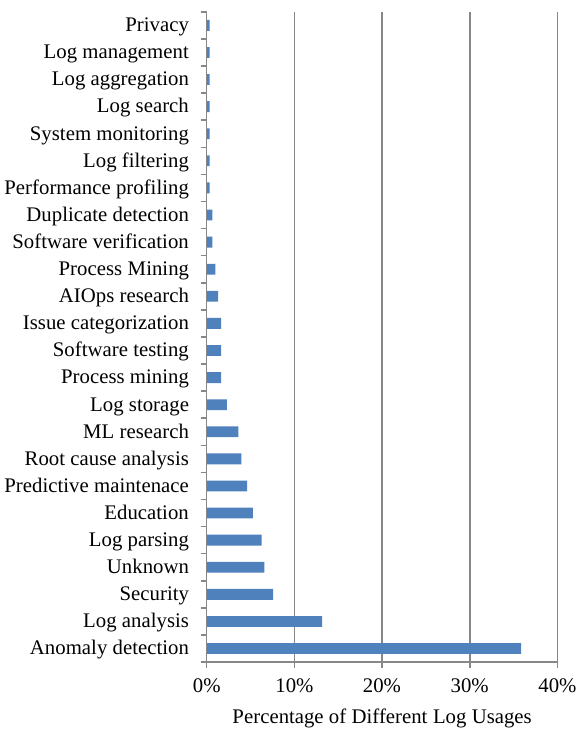}
\caption{Summary of Tentative Industry Adoption of Loghub}
\label{fig:industry}
\end{figure}

The loghub datasets have been made available for five years. During this time, we conducted a survey via Zenodo, a dataset hosting website, to gather information regarding the organization and usage scenarios of dataset downloading requests. Since its release, loghub has attracted the attention of not only large companies such as IBM, Microsoft, Huawei, Nvidia, MasterCard, Adobe, BMW, and Samsung, but also some startup companies focusing on building log analysis products, including Elastic.co, Splunk, Rapid7, Element AI, White Ops, Unomaly.com, and Ascend.io\footnote{See \url{https://github.com/logpai/loghub/wiki/Loghub-download-list}.}. Many universities have also requested the loghub dataset. To date, loghub has been downloaded 90,000+ times by more than 450 organizations from both industry (35\%) and academia (65\%). After analyzing the information collected from the data requests received, we manually categorize the log usage scenarios and present the distribution in Figure~\ref{fig:industry}. It is worth noting that due to the limited information provided by users, we could only provide a rough categorization for them. We denote it as an unknown category if the user input is not clear. We can see that loghub datasets can potentially facilitate 23 different categories of research and education purposes. The top 5 usage scenarios are anomaly detection, log analysis, security, log parsing, and education. In particular, log analysis may cover the spectrum of some other log related tasks, but this has not been clearly specified by users. Education indicates the use cases about course projects and thesis projects. Here, we do not intend to expand all usage scenarios shown in the figure, but the variety of them have already confirmed the practical importance of the loghub datasets. 

In the following, we present the details of four common usage scenarios that have been widely studied in the literature and describe how loghub can be used in these tasks, including log parsing, log compression, anomaly detection, and duplicate issue identification.

\subsection{Log Parsing}
Most of the AI-based log analysis approaches require structured input data, such as a list of system events with event IDs or a matrix. However, software logs are often unstructured texts, containing several fields and natural language descriptions written by developers. Thus, log parsing~\cite{zhu2018tools} is a crucial step in AI-driven log analytics that transforms unstructured log messages into structured system events. 

However, as the volume of logs increases rapidly, traditional parsing approaches that largely rely on manual parsing rules construction becomes labor-intensive and inefficient. To address this problem, recent research has proposed various data-driven log parsers \cite{SLCT, nagappan2010abstracting, vaarandi2015logcluster, LKE, LogSig, hamooni2016logmine, SHISO, shima2016length, jiang2008abstracting, IPLoM, He17ICWS, spellICDM16, wang2022spine, xu2023hue, huo2023semparser}, which automatically label an unstructured log message with corresponding system event ID. Typically, log parsing is modeled as a clustering problem, where log messages describing the same system event should be clustered into the same group. The common tokens in all the log messages in the same group is regarded as the system event or event template. The research problem of log parsing is how to accurately and efficiently separate the unstructured log messages into different clusters by designing similarity metrics for log messages and novel clustering approaches. By clustering log messages into groups, log parsers can summarize the corresponding system events and match each log message with an event ID. The structured logs (i.e., log messages with event ID) could be easily transformed into a matrix or directly utilized by log analysis algorithms.

To evaluate the accuracy and efficiency of log parsing approaches, we need: (1) a large volume of logs and (2) logs generated by a variety of systems. Loghub contains 19 log datasets collected from 6 categories of systems. Besides, all the logs amount to over 77 GB. Thus, datasets in loghub can be employed in the experiments to evaluate the parsing accuracy and efficiency of different log parsing approaches~\cite{zhu2018tools}.

\subsection{Log Compression}
Logs can be used in various system maintenance tasks, and thus they often need to be stored for a long time (e.g., a year or more) in practice. As the the explosion of log size in recent years, archiving system logs is consuming a large amount of storage space, which leads to high cost of electrical power. General compression approaches do not work well on log compression because they do not consider the inherent structure of log messages. To achieve higher compression ratio, a new line of research~\cite{MLC-9, MLC-18, MLC-5, MLC-11, MLC} has proposed compression approaches specialized for log data.

Log compression can be modeled as a frequent pattern mining problem. Existing approaches focus on finding inherent structure information of log messages (e.g., repetitive text). In particular, these approaches provide different strategies to detect repetitive text, such as utilizing the common format of logs generated by a specific system~\cite{MLC-5}. The research problem of log compression is how to achieve efficient and lossless compression with high compression rate.

To evaluate the accuracy and efficiency of log compression approaches, similar to the evaluation of log parsers, we need a large volume of logs collected from diverse systems. Thus, all the datasets in loghub can facilitate the evaluation of log compression approaches, as demonstrated in~\cite{liu2019logzip}.

\subsection{Anomaly Detection}
Modern systems have become large-scale in size and complex in structure. An increasing number of these systems are expected to run on a $24\times7$ basis serving millions of users globally. Any non-trivial downtime of them could lead to enormous revenue loss~\cite{anomaly1, anomaly2}. Thus, to enhance the reliability of modern systems, a line of recent research~\cite{Xu_sosp_2009, lou2010mining, bodik2010fingerprinting, ll-25, slheISSRE16, lin2016log, du2017deeplog, chen2021experience, liu2023scalable, huo2023evlog} has focused on log-based anomaly detection approaches that report potential abnormal system behaviors by analyzing system runtime logs.

The anomaly detection problem is usually modeled as a binary classification problem. The input is a list of structured system events or a matrix, while the output is a list of labels indicating whether an instance (e.g., an event or a time period) is abnormal. There are mainly two categories of log-based anomaly detection approaches: \textit{unsupervised}~\cite{Xu_sosp_2009, lou2010mining, lin2016log} and \textit{supervised}~\cite{bodik2010fingerprinting, ll-25, du2017deeplog}. The research problem of log-based anomaly detection is how to accurately detect the anomalies based on system logs. Towards this end, F-measure (i.e., F1 score)~\cite{IRbook08}, a commonly-used evaluation metric for classification algorithms, is employed. 

To evaluate the accuracy of log-based anomaly detection approaches, we need log datasets that contain anomaly labels for instances (e.g., whether a time period is regarded as anomaly). Loghub contains 6 labeled log datasets, which can be used in the experiments to evaluate the accuracy of diverse anomaly detection approaches~\cite{slheISSRE16}.

\subsection{Duplicate Issues Identification}
To enhance system reliability, one important task for developers is to handle user-reported operational issues efficiently. An operational issue is a system problem reported by users. When a user of Amazon EC2 finds that her node becomes extremely slow, she will report the node slowness as an issue to Amazon. To handle an operational issue, developers need to mainly inspect the runtime logs to understand the system operations, which is time-consuming. Thus, to facilitate the issue handling process, recent research has proposed duplicate issues identification techniques~\cite{ding2014mining, lim2014identifying, rakha2018revisiting, liu2023incident} to alleviate unnecessary manual effort. 

The duplicate issue identification problem can be modeled as a clustering problem, while the duplicate issues are clustered into the same group based on the corresponding log messages. If the log messages of two issues demonstrate similar patterns (e.g., occurrence frequency, order), the two issue will be clustered in to the same group. The research problem of log-based duplicate issues identification is how to accurately separate the log messages (i.e., log sequences) of different issues into clusters.

To evaluate the accuracy of log-based duplicate issues identification approaches, log datasets that contain the issue categories are needed. Loghub contains 3 labeled log datasets (i.e., HDFS-v1, Hadoop, Android-v2) that provide this information. Thus, loghub could be used for this task.

%% file: Sections/casestudy_parsing.tex
\section{Benchmarking on Loghub Datasets}\label{sec:benchmarks}
In this section, we demonstrate the use of loghub dataset via benchmarking typical log analysis tasks including log parsing, log compression, and log-based anomaly detection. From the results of this benchmarking, we derive critical unresolved questions and challenges inherent to each task. Our aspiration is that the academic community may leverage the insights drawn from the large-scale dataset, loghub, thereby fostering further advancements in the field.

\subsection{Benchmarking for Log Parsing}
In the following, we describe a case study of benchmarking existing log parsing algorithms using loghub.

\begin{table}[t]
\centering{}
\caption{Summary of log parsing tools.}
\includegraphics[scale=0.65]{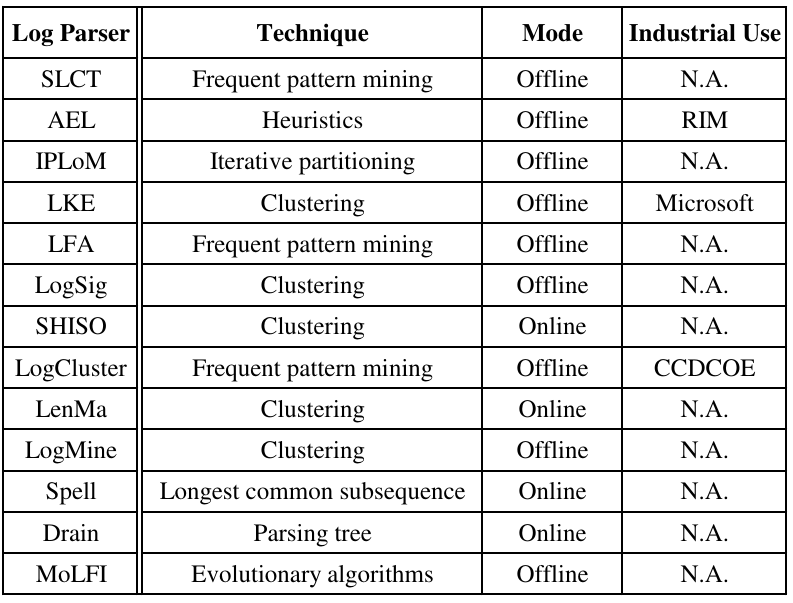}
\label{tab:parsers}
\end{table}

\begin{table*}[t]
\centering{}
\caption{Accuracy of existing log parsing approaches.}
\includegraphics[scale=1]{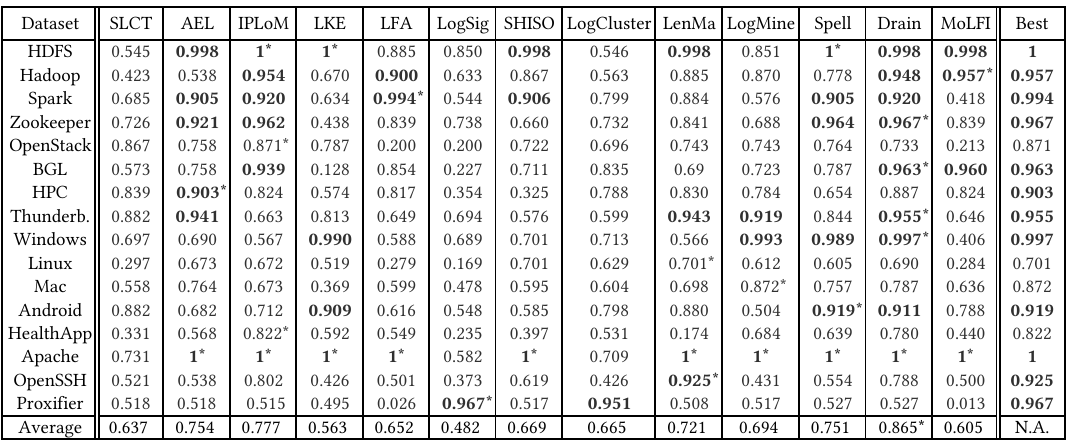}
\label{tab:parsing_acc}
\end{table*}

\subsubsection{Existing Log Parsing Algorithms}
We have evaluated 13 log parsing algorithms as shown in Table~\ref{tab:parsers}, which can be categorized into four types: \textit{frequent pattern-based}, \textit{clustering-based}, \textit{heuristics-based} methods and \textit{others}.

Typical \textit{frequent pattern-based} methods include SLCT~\cite{SLCT}, LFA~\cite{nagappan2010abstracting} and LogCluster~\cite{vaarandi2015logcluster}. They require the entire set of log messages ready before parsing, i.e., offline mode. They have a similar rationale that first identify frequent patterns (e.g., tokens or token-position pairs) then group all log messages based on these patterns to clusters. The templates are finally extracted from each cluster.
\textit{Clustering-based} methods include LKE~\cite{LKE}, LogSig~\cite{LogSig}, SHISO~\cite{SHISO}, LenMa~\cite{shima2016length} and LogMine~\cite{hamooni2016logmine}. These methods rely on a core clustering algorithm (e.g., hierarchical clustering) to group log messages to clusters, then extract a template from each cluster. Among these clustering-based methods, SHISO and Lenma have an online mode that can process each log message without seeing the entire log set. LKE, LogSig and LogMine only have an offline mode.
\textit{Heuristics-based} methods include AEL~\cite{jiang2008abstracting}, IPLoM~\cite{IPLoM} and Drain~\cite{Drain}. Specifically, AEL sorts log messages into groups by checking the frequency of static and variable tokens. IPLoM uses a step-by-step division method to group messages based on length, token location, and mapping connection. Drain uses a fixed-depth tree model to display log messages and quickly extracts common patterns. These methods use log features effectively and often yield good results.
\textit{Other} methods include Spell~\cite{spellICDM16} and MoLFI~\cite{messaoudi2018search}. Spell applies an algorithm based on the longest common subsequence for continuous log parsing. MoLFI formulates log parsing as a problem of multi-objective optimization, and resolves it through evolutionary algorithms.

\subsubsection{Benchmarking on Loghub}

To evaluate the accuracy of different log parsing algorithms. We define the metric of parsing accuracy (PA) as follows:
\begin{align}
    PA = \frac{\#\ of\ corrected\ parsed\ logs}{\#\ of\ total\ logs}
\end{align}
After parsing, every log message transforms into an event template, and each template relates to a cluster of log messages sharing the same template. 
A log message is considered correctly parsed if and only if its event template matches the same cluster of log messages as the groudtruth does.
For instance, parsing a log sequence [E1, E2, E2] to [E1, E4, E5] results in a Parsing Accuracy (PA) of 1/3, as the second and third messages are not grouped similarly.

We benchmark these log parsing algorithms in Table~\ref{tab:parsers} using the datasets in Loghub, and the experimental results are shown in Table~\ref{tab:parsing_acc}.
For each dataset, the best accuracy is highlighted using an asterisk “*” and shown in the column ``Best'', and the PA higher than 0.9 are marked as boldface.
We can summarize the following observations: 
(1) Over 90\% accuracy is achieved by at least one log parser on most datasets, with 8 out of 13 log parsers showing the best accuracy on at least two datasets. Some parsers can even handle HDFS and Apache datasets with 100\% accuracy due to their simpler event templates.
(2) Despite these successes, complex log types like OpenStack, Linux, Mac, and HealthApp still present challenges due to their intricate structures and numerous event templates, such as the 341 templates in Mac logs. This calls for further improvements in parsing these complex log data.
(3) Regarding the overall effectiveness of log parsers, Drain stands out with the highest average accuracy across different datasets, showing high precision on 9 out of 16 datasets. IPLoM, AEL, and Spell also perform well, maintaining high accuracy on 6 datasets. Conversely, LogSig, LFA, MoLFI, and LKE have the lowest average accuracy.

\begin{table*}[t]
\centering{}
\caption{Compression effectiveness of existing log compression approaches.}
\includegraphics[scale=0.69]{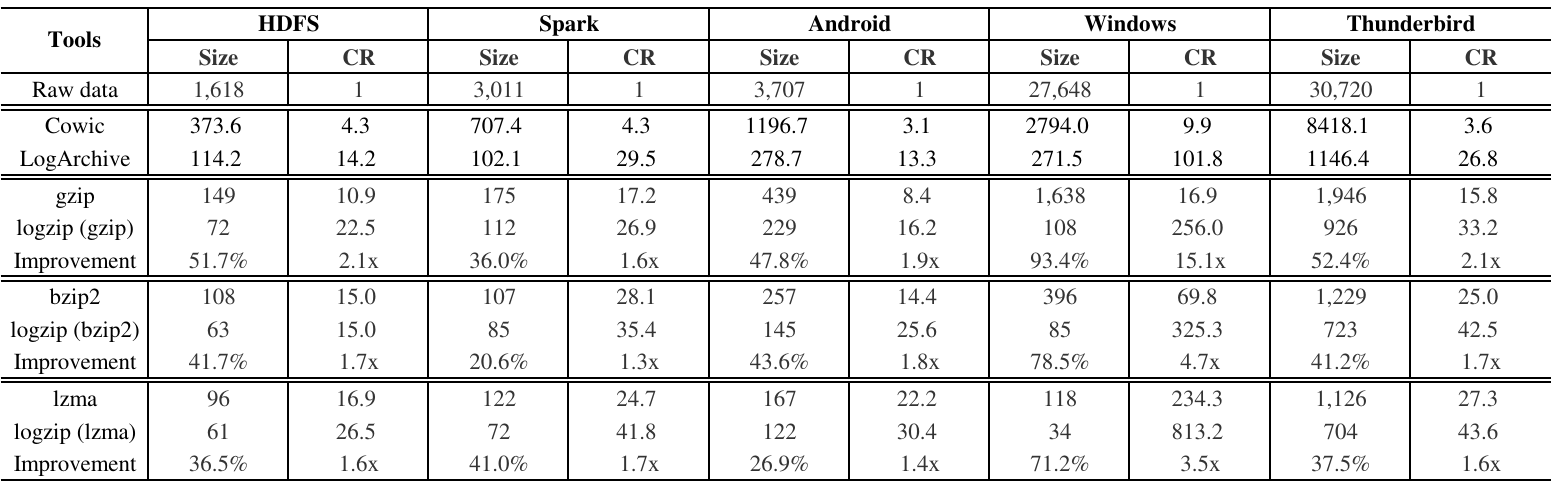}
\label{tab:compression_acc}
\vspace{-10pt}
\end{table*}

\subsubsection{Remaining Questions and Challenges}
Based on the benchmarking results, the following questions and challenges can be summarized:
(1) No single log parser can effectively handle all types of datasets. This requires the development of a more generalized log parsing algorithm capable of handling diverse types of log data.
(2) Dealing with complex datasets containing a substantial number of templates poses a significant challenge for most existing methods. A practical log parser should possess the ability to accurately handle such logs.
(3) Existing log parsers can only differentiate between log templates and log parameters. It would potentially enhance the diagnostic procedures of on-site engineers if log parameters could be classified into various fine-grained types, such as status codes and block IDs.
(4) The majority of log parsers rely on identifying frequent patterns to determine templates. However, log message occurrence distributions in practical scenarios can be diverse, resulting in instances where certain log messages appear only a few times. This leads to decreased parsing accuracy.

%% file: Sections/casestudy_compression.tex
\subsection{Benchmarking for Log Compression}
In this section, we demonstrate a case study of bechmarking existing log compression algorithms using loghub.

\subsubsection{Existing Log Compression Algorithms}
We evaluate 6 compression tools which can be categorized to \textit{log-specific compression tools} including Cowic~\cite{MLC-11}, LogArchive~\cite{logarchive} and Logzip~\cite{liu2019logzip} and \textit{general compression} tools including gzip, bzip2 and lzma.
Note that Logzip should be utilized in conjunction with general compression tools (called compression kernels of Logzip), resulting in three variants: Logzip (gzip), Logzip (bzip2), and Logzip (lzma).
Specifically, log-specific compression tools are specifically designed to compress log data by exploiting the inherent structures present in logs. This specialized approach enables these tools to achieve improved compression ratios for log files.
On the other hand, general compression tools are designed to compress various types of files. These tools primarily identify redundant sequences of bytes within the compression objects and employ Huffman Coding techniques to minimize the required storage space.

\subsubsection{Benchmarking on Loghub}
To evaluate the effectiveness of the tools for log compression, we use the metric named compression ratio defined as follows:
$$ CR = \frac{Original\ File\ Size}{Compressed\ File \ Size},$$
where the compressed file size is obtained after applying a compression tool. A smaller compressed file size can result in a higher CR, indicating a more effective compression tool.

We utilize five large-scale datasets within Loghub to benchmark these compression tools in terms of CR. The experimental results are shown in Table~\ref{tab:compression_acc}. We can make the following observations:
(1) Among the general compression tools (lzma, bzip2, and gzip), lzma is the most effective on most datasets, followed by bzip2, while gzip performs the worst.
(2) Two algorithms designed specifically for log data, LogArchive and Cowic, offer varying performance. LogArchive achieves a higher CR than gzip but is less effective than bzip2 and lzma, while Cowic performs worse than gzip because it prioritizes quick queries instead of a high CR.
(3) Logzip, equipped with different compression kernels, outperforms these methods, with the compressed size determined by the kernel's effectiveness. It achieves higher CR on all five datasets, with an average CR of 4.56x and a maximum of 15.1x over gzip, resulting in significant storage savings. Similar results are observed with other compression kernels.

\subsubsection{Remaining Questions and Challenges}
According to the benchmarking results, we can summarize the following remaining questions and challenges.
(1) The effectiveness of log-specific compression tools is impacted by the number of templates within the target log data. For example, the CR of different datasets varies a lot. 
Hence, it is crucial to conduct an in-depth study into how template distribution influences the performance of log-specific compression tools.
(2) Current log-specific compression tools primarily concentrate on optimizing the CR, often at the expense of search efficiency. Practical applications frequently require on-site engineers to locate specific templates or keywords within compressed log files. Consequently, the issue of simultaneously achieving high CR and maintaining robust search efficiency remains an open question in the field.
(3) The resource consumption of existing compression tools is still understudied. Log compression algorithms can be deployed on nodes with limited computational capacity or memory in real-world scenarios. Given their function as background processes, these algorithms must be sufficiently lightweight to prevent excessive overhead imposition on the host machine, an aspect yet to be thoroughly investigated.

%% file: Sections/casestudy_ad.tex
\subsection{Benchmarking for Anomaly Detection}\label{sec:casestudy}

In this section, we demonstrate the usage of loghub by a case study on benchmarking existing log-based anomaly detection approaches. 

\subsubsection{Existing Log-based Anomaly Detection Approaches}

\begin{table}[t]
\centering{}
\caption{Summary of Log-based Anomaly Detection.}
\includegraphics[scale=0.5]{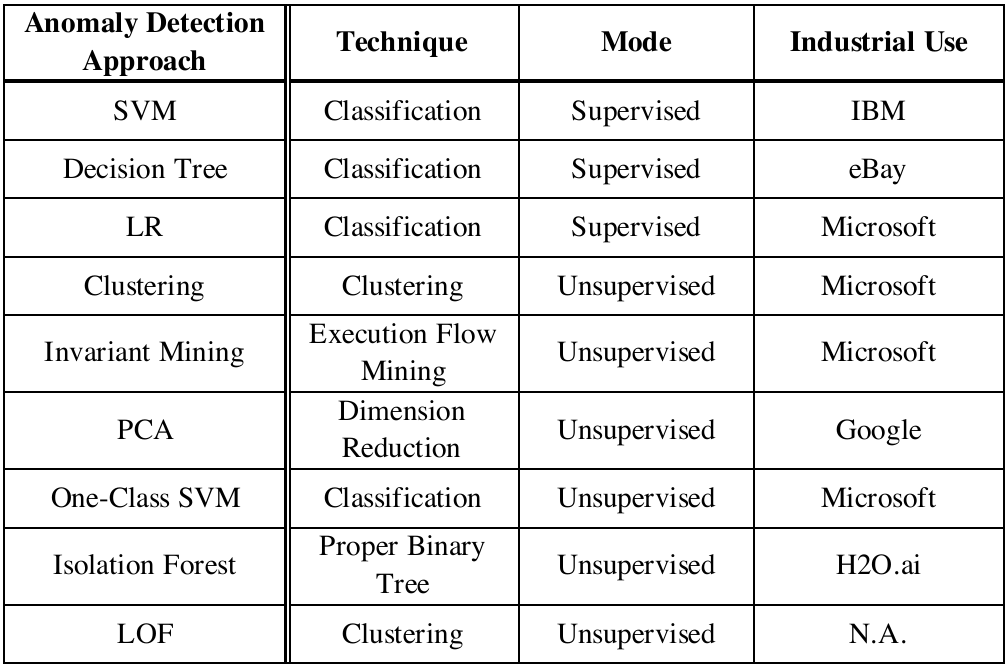}
\label{tab:loglizer}
\vspace{-10pt}
\end{table}

We have evaluated the performance of 9 log-based anomaly detection approaches, which are illustrated in Table~\ref{tab:loglizer}. There are mainly two categories of anomaly detection approaches: \textit{supervised} and \textit{unsupervised}. 

\textit{Supervised} approaches (Decision Tree~\cite{chen2004failure}, SVM~\cite{liang2007failure}, and LR~\cite{bodik2010fingerprinting}) require training data that contain labels indicating whether an instance is an anomaly. A classifier is trained based on the labeled data for anomaly detection. Supervised approaches are used when there are decent amount of both normal and abnormal labeled data. \textit{Unsupervised} approaches are based on different techniques, including classification (LR~\cite{scholkopf2001estimating}), isolation via proper binary tree (Isolation Forest~\cite{liu2008isolation}), dimension reduction (PCA~\cite{Xu_sosp_2009}), execution flow mining (Invariant Mining~\cite{lou2010mining}), and clustering (LOF~\cite{breunig2000lof} and Clustering~\cite{lin2016log}). The core idea of unsupervised approach is to learn the common patterns in logs or log sequences, and report instances the deviating instances as anomalies. In practice, labeled data are often lacking, because (1) anomalies rare occur in real-world systems and (2) data labeling is label-intensive and time-consuming. Thus, unsupervised methods are more applicable in real-world production environment. 

\subsubsection{Benchmarking on Loghub}
To evaluate the accuracy of anomaly detection, we use \textit{precision}, \textit{recall}, and \textit{F-measure} (i.e., F1 score), which are the most commonly-used metrics~\cite{slheISSRE16}. 









All the anomaly detection approaches are evaluated on the labeled HDFS dataset. This dataset records the system operations on different HDFS blocks. After log parsing and some postprocessing, we can obtain the input: a \textit{block-ID}-by-\textit{event count} matrix. Each row of the matrix represents the system operations on a specific block, while each column represents the frequency of occurrence of a system event during runtime.

\begin{figure}[h]
\centering{} 
\includegraphics[scale=0.95]{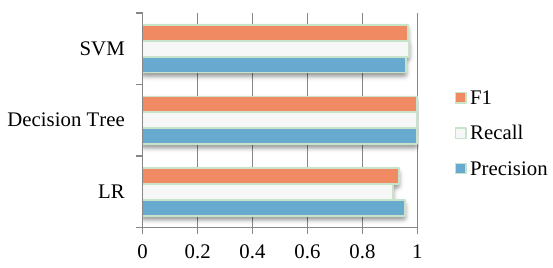}\hspace{-13pt}
\vspace{-2ex}
\caption{Accuracy of Supervised Anomaly Detection Approaches}
\label{fig:supervisedresult}
\end{figure}

\begin{figure}[h]
\centering{} 
\includegraphics[scale=0.9]{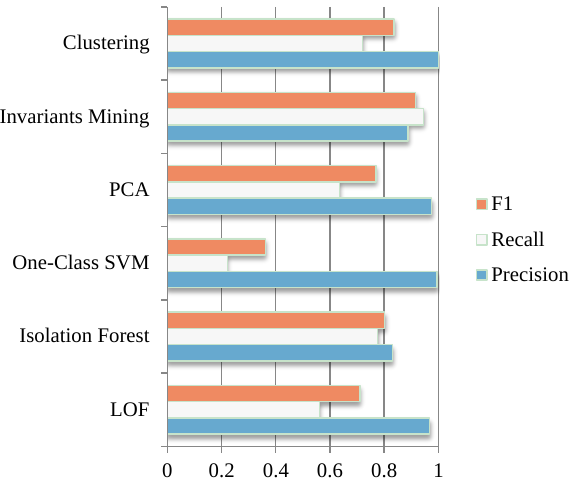}
\caption{Accuracy of Unsupervised Anomaly Detection Approaches}
\label{fig:unsupervisedresult}
\vspace{-11pt}
\end{figure}

The experimental results are illustrated in Figure~\ref{fig:supervisedresult} and Figure~\ref{fig:unsupervisedresult}. \textit{Decision Tree} obtains the highest recall (0.99), F-measure (0.99), and the second-highest precision (0.99). The supervised approaches achieve better results. This is because the supervised approaches are trained on labeled data. Additionally, among the unsupervised approaches, This is because the supervised approaches are trained on labeled data. Additionally, among the unsupervised approaches, \textit{Invariants Mining} approach has the best accuracy. \textit{Clustering} obtains higher precision (1.00) than Invariants Mining. However, its recall (0.72) is much lower. Although \textit{One-Class SVM} obtain high precision (0.99), its recall is too low (0.22), leading to low F-measure. This is because One-Class SVM is too conservative on reporting anomalies. 

In additional to the labeled HDFS dataset used to evaluate anoamly detection method in this section, loghub also provides other 4 labeled datasets. Thus, in practice, developers could evaluate an anomaly detection method on different datasets and choose the most suitable one.

\subsubsection{Remaining Questions and Challenges}
After benchmarking existing anomaly detection approaches on loghub, we find some remaining questions and challenges, which will be discussed in this session. 
(1) The unsupervised approaches are not as accurate as the supervised approaches. Thus, an accurate unsupervised anomaly detection approach is highly in demand. 
(2) In practice, the number of anomalies is much less than that of normal instances. In industry, a system may only encounter 1 anomaly in a year, which makes supervised approaches ineffective. Thus, how to design an anomaly detection approach that does not require historical abnormal instances remains an important and challenging problem. 
(3) Current anomaly detection approaches are all log sequences-based, which provide limited help to developers on further diagnosis, such as root cause analysis. 
(4) Most of the existing approaches will only report whether an instance is an anomaly without other information. Thus, there is a lack of visualization tools that help developers understand the reported anomalies.
(5) Existing approaches mainly focus on improving accuracy for anomaly detection. However, a practical solution should be efficient enough to process a large number of log data in the runtime.
(6) In real-world scenarios, log data evolves alongside software development. This phenomenon, known as concept drift, poses a challenge to existing methods in maintaining satisfactory performance.

%% file: Sections/relatedwork.tex
\section{Related Work}\label{sec:relatedwork}

\textbf{Logging Practice}. In practice, there is a lack of rigorous guide and specifications on developer logging behaviors. To address this problem, a line of recent empirical research has focused on studying the logging practice of high-quality software~\cite{barik2016bones, chen2017characterizing, fu2014developers, li17ESEa, pecchia2015industry, yuan2012characterizing, yuan2012conservative}. Additionally, AI-based logging decisions have also been widely studied in recent years with a focus on "where to log" and "what to log". \textit{Where-to-log}: The work~\cite{zhu15learningToLog} proposed a "learning to log" framework, which provides guidance on whether developers should put logging statement in a code snippet. \textit{What-to-log}: Li et al.~\cite{li17ESEb} developed an log verbosity level suggestion technique based on ordinal regression model. He et al. \cite{he2018characterizing} characterizes the natural language descriptions in logging statements and explore the potential of automated description text generation for logging statements. AI-based strategic logging practices is an important component in AI-driven log analytics, where different practice will lead to different software logs at runtime.

\textbf{Log Compression}. Different from general natural language text, software logs have specific inherent structure (e.g., many logs are printed by the same logging statement). Thus, to achieve better compression rate, compression approaches specialized for log files have been well studied~\cite{MLC-9, MLC-18, MLC-5, MLC-11, MLC}. For example, Comprehensive Log Compression (CLC)~\cite{MLC-9} and Differentiated Semantic Log Compression (DSLC)~\cite{MLC-18} identify repetitive items by domain knowledge.
Logzip~\cite{liu2019logzip} extracts redundant log templates using log parsing technologies, which improves over existing compression tools such as gzip.
The work \cite{wei2021feasibility}  further improves logzip with a more efficient implementation for compressing huge logs in cloud systems.
Beyond compression, CLP~\cite{rodrigues2021clp} and Loggrep~\cite{wei2023loggrep} achieve more search efficiency on compressed log data without decompression.
Logreducer~\cite{yu2023logreducer} detects log hot spot in the runtime to avoid saving massive and redundant logs.
All the datasets in loghub can facilitate the evaluation of log compression approaches.

\textbf{Log Parsing}. Most automated and effective log analysis techniques require structured data as input. Therefore, log parsing is crucial in AI-driven log analytics, which transforms unstructured log messages into structured system events. In recent years, log parsers based on various techniques have been proposed. (1) \textit{Frequent pattern mining}: SLCT~\cite{SLCT}, LFA~\cite{nagappan2010abstracting}, and LogCluster~\cite{vaarandi2015logcluster} regard log event templates as a set of constant tokens that occur frequently in log. (2) \textit{Clustering}: a line of log parsing studies model it as a clustering problem and design specialized clustering algorithms accordingly. Typical parsers in this category include LKE~\cite{LKE}, LogSig~\cite{LogSig}, LogMine~\cite{hamooni2016logmine}, SHISO~\cite{SHISO}, and LenMa~\cite{shima2016length}. (3) \textit{Heuristics}: AEL~\cite{jiang2008abstracting}, IPLoM~\cite{IPLoM}, Drain~\cite{He17ICWS} and SPINE~\cite{spine} parse log messages by heuristic rules inspired by unique characteristics of software logs. It is worth noting that Drain~\cite{He17ICWS} has been a widely adopted solution in industry, such as IBM. (4) \textit{Deep Learning}: UniParser~\cite{uniparser}. (5) \textit{Others}: Some other methods exist, such as Spell~\cite{spellICDM16} that is based on longest common subsequence. All the datasets in loghub can be employed to evaluate log parsing methods.  


\textbf{Log Analysis}.
Log analysis has been studied for decades to facilitate effective and efficient system maintenance~\cite{cheng2023logai}. Typical log analysis tasks include anomaly detection~\cite{Xu_sosp_2009, ll-25, slheISSRE16, du2017deeplog}, duplicate issue identification~\cite{ding2014mining, lim2014identifying, rakha2018revisiting}, incident diagnosis~\cite{onion}, usage statistics analysis~\cite{ll-31}, and program verification~\cite{beschastnikh2011leveraging, verification}, most of which design or adopt AI algorithms. For example, Xu et al.~\cite{Xu_sosp_2009} employs principal component analysis (PCA), which is a dimension reduction algorithm, to detect potential anomalies in large-scale distributed systems. Du et al.~\cite{du2017deeplog} proposed an anomaly detection and diagnosis approach based on a deep neural network model utilizing Long Short-Term Memory (LSTM). Shang et al.~\cite{verification} design a clustering algorithm to verify the deployment of big data applications. The 5 labeled datasets in loghub can be used to evaluate various log analysis methods.

%% file: Sections/conclusion.tex
\section{Conclusion and Future Work}\label{sec:conclusion}

This paper describes {loghub}, a large collection of log datasets for AI-driven log analytics. Loghub contains 19 log datasets where all the logs amount to over 77 GB. We present both common usage scenarios and benchmarking results for typical log analysis tasks including log parsing, log compression, and log-based anomaly detection.
Based on these benchmarking results, we also discuss open challenges and future directions.
We envision loghub acting as an open benchmarking system for AI-driven log analysis to benefit researchers and practitioners from academia and industry. As part of our future work, we plan to collect more large-scale and more abundant log datasets to fill the gap between research and practice. Additionally, we aim to build a benchmarking leaderboard based on loghub to assess log analysis models and tasks.

\section{Acknowledgement}
The work described in this paper was supported by the Research Grants Council of the Hong Kong Special Administrative Region, China (No. CUHK 14206921 of the General Research Fund).

%% file: issre23.bbl
\begin{thebibliography}{10}
\providecommand{\url}[1]{#1}
\csname url@samestyle\endcsname
\providecommand{\newblock}{\relax}
\providecommand{\bibinfo}[2]{#2}
\providecommand{\BIBentrySTDinterwordspacing}{\spaceskip=0pt\relax}
\providecommand{\BIBentryALTinterwordstretchfactor}{4}
\providecommand{\BIBentryALTinterwordspacing}{\spaceskip=\fontdimen2\font plus
\BIBentryALTinterwordstretchfactor\fontdimen3\font minus
  \fontdimen4\font\relax}
\providecommand{\BIBforeignlanguage}[2]{{%
\expandafter\ifx\csname l@#1\endcsname\relax
\typeout{** WARNING: IEEEtranS.bst: No hyphenation pattern has been}%
\typeout{** loaded for the language `#1'. Using the pattern for}%
\typeout{** the default language instead.}%
\else
\language=\csname l@#1\endcsname
\fi
#2}}
\providecommand{\BIBdecl}{\relax}
\BIBdecl

\bibitem{barik2016bones}
T.~Barik, R.~DeLine, S.~Drucker, and D.~Fisher, ``The bones of the system: A
  case study of logging and telemetry at microsoft,'' in \emph{2016 IEEE/ACM
  38th International Conference on Software Engineering Companion
  (ICSE-C)}.\hskip 1em plus 0.5em minus 0.4em\relax IEEE, 2016, pp. 92--101.

\bibitem{beschastnikh2011leveraging}
I.~Beschastnikh, Y.~Brun, S.~Schneider, M.~Sloan, and M.~D. Ernst, ``Leveraging
  existing instrumentation to automatically infer invariant-constrained
  models,'' in \emph{Proceedings of the 19th ACM SIGSOFT symposium and the 13th
  European conference on Foundations of software engineering (FSE)}.\hskip 1em
  plus 0.5em minus 0.4em\relax ACM, 2011, pp. 267--277.

\bibitem{bodik2010fingerprinting}
P.~Bodik, M.~Goldszmidt, A.~Fox, D.~B. Woodard, and H.~Andersen,
  ``Fingerprinting the datacenter: automated classification of performance
  crises,'' in \emph{Proceedings of the 5th European Conference on Computer
  Systems (EuroSys)}.\hskip 1em plus 0.5em minus 0.4em\relax ACM, 2010, pp.
  111--124.

\bibitem{breunig2000lof}
M.~M. Breunig, H.-P. Kriegel, R.~T. Ng, and J.~Sander, ``Lof: identifying
  density-based local outliers,'' in \emph{ACM sigmod record}, vol.~29,
  no.~2.\hskip 1em plus 0.5em minus 0.4em\relax ACM, 2000, pp. 93--104.

\bibitem{chen2017characterizing}
B.~Chen and Z.~M.~J. Jiang, ``Characterizing logging practices in java-based
  open source software projects--a replication study in apache software
  foundation,'' \emph{Empirical Software Engineering}, vol.~22, no.~1, pp.
  330--374, 2017.

\bibitem{chen2004failure}
M.~Chen, A.~X. Zheng, J.~Lloyd, M.~I. Jordan, and E.~Brewer, ``Failure
  diagnosis using decision trees,'' in \emph{International Conference on
  Autonomic Computing (ICAC)}.\hskip 1em plus 0.5em minus 0.4em\relax IEEE,
  2004, pp. 36--43.

\bibitem{chen2021experience}
Z.~Chen, J.~Liu, W.~Gu, Y.~Su, and M.~R. Lyu, ``Experience report: Deep
  learning-based system log analysis for anomaly detection,'' \emph{arXiv
  preprint arXiv:2107.05908}, 2021.

\bibitem{cheng2023logai}
Q.~Cheng, A.~Saha, W.~Yang, C.~Liu, D.~Sahoo, and S.~Hoi, ``Logai: A library
  for log analytics and intelligence,'' \emph{arXiv preprint arXiv:2301.13415},
  2023.

\bibitem{logarchive}
R.~Christensen and F.~Li, ``Adaptive log compression for massive log data,'' in
  \emph{Proceedings of the {ACM} {SIGMOD} International Conference on
  Management of Data (SIGMOD)}, 2013, pp. 1283--1284.

\bibitem{securitylogproject}
\BIBentryALTinterwordspacing
A.~Chuvakin. (2019) Public security log sharing site project. [Online].
  Available: \url{http://log-sharing.dreamhosters.com/}
\BIBentrySTDinterwordspacing

\bibitem{cloudlab}
\BIBentryALTinterwordspacing
CloudLab. (2019) Cloudlab. [Online]. Available: \url{https://cloudlab.us/}
\BIBentrySTDinterwordspacing

\bibitem{MLC-5}
S.~Deorowicz and S.~Grabowski, ``Sub-atomic field processing for improved web
  log compression,'' in \emph{Modern Problems of Radio Engineering,
  Telecommunications and Computer Science, 2008 Proceedings of International
  Conference on}.\hskip 1em plus 0.5em minus 0.4em\relax IEEE, 2008, pp.
  551--556.

\bibitem{ding2014mining}
R.~Ding, Q.~Fu, J.~G. Lou, Q.~Lin, D.~Zhang, and T.~Xie, ``Mining historical
  issue repositories to heal large-scale online service systems,'' in
  \emph{44th Annual IEEE/IFIP International Conference on Dependable Systems
  and Networks (DSN)}.\hskip 1em plus 0.5em minus 0.4em\relax IEEE, 2014, pp.
  311--322.

\bibitem{spellICDM16}
M.~Du and F.~Li, ``Spell: Streaming parsing of system event logs,'' in
  \emph{ICDM'16 Proc. of the 16th International Conference on Data Mining},
  2016.

\bibitem{du2017deeplog}
M.~Du, F.~Li, G.~Zheng, and V.~Srikumar, ``Deeplog: Anomaly detection and
  diagnosis from system logs through deep learning,'' in \emph{Proceedings of
  the 2017 ACM SIGSAC Conference on Computer and Communications
  Security}.\hskip 1em plus 0.5em minus 0.4em\relax ACM, 2017, pp. 1285--1298.

\bibitem{MLC}
B.~Feng, C.~Wu, and J.~Li, ``{MLC:} an efficient multi-level log compression
  method for cloud backup systems,'' in \emph{2016 {IEEE}
  Trustcom/BigDataSE/ISPA, Tianjin, China, August 23-26, 2016}, 2016, pp.
  1358--1365.

\bibitem{qfuICSE14q}
Q.~Fu, J.~Zhu, W.~Hu, J.~Lou, R.~Ding, Q.~Lin, D.~Zhang, and T.~Xie, ``Where do
  developers log? an empirical study on logging practices in industry,'' in
  \emph{ICSE'14: Companion Proc. of the 36th International Conference on
  Software Engineering}, 2014, pp. 24--33.

\bibitem{ll-25}
Q.~Fu, J.~Lou, Y.~Wang, and J.~Li, ``Execution anomaly detection in distributed
  systems through unstructured log analysis,'' in \emph{{ICDM} 2009, The Ninth
  {IEEE} International Conference on Data Mining, Miami, Florida, USA, 6-9
  December 2009}, 2009, pp. 149--158.

\bibitem{LKE}
------, ``Execution anomaly detection in distributed systems through
  unstructured log analysis,'' in \emph{{ICDM} 2009, The Ninth {IEEE}
  International Conference on Data Mining, Miami, Florida, USA, 6-9 December
  2009}, 2009, pp. 149--158.

\bibitem{fu2014developers}
Q.~Fu, J.~Zhu, W.~Hu, J.-G. Lou, R.~Ding, Q.~Lin, D.~Zhang, and T.~Xie, ``Where
  do developers log? an empirical study on logging practices in industry,'' in
  \emph{Companion Proceedings of the 36th International Conference on Software
  Engineering}.\hskip 1em plus 0.5em minus 0.4em\relax ACM, 2014, pp. 24--33.

\bibitem{hamooni2016logmine}
H.~Hamooni, B.~Debnath, J.~Xu, H.~Zhang, G.~Jiang, and A.~Mueen, ``Logmine:
  Fast pattern recognition for log analytics,'' in \emph{Proceedings of the
  25th ACM International on Conference on Information and Knowledge
  Management}.\hskip 1em plus 0.5em minus 0.4em\relax ACM, 2016, pp.
  1573--1582.

\bibitem{MLC-9}
K.~H{\"{a}}t{\"{o}}nen, J.~Boulicaut, M.~Klemettinen, M.~Miettinen, and
  C.~Masson, ``Comprehensive log compression with frequent patterns,'' in
  \emph{Data Warehousing and Knowledge Discovery, 5th International Conference,
  DaWaK 2003, Prague, Czech Republic, September 3-5,2003, Proceedings}, 2003,
  pp. 360--370.

\bibitem{He17ICWS}
P.~He, J.~Zhu, Z.~Zheng, and M.~R. Lyu, ``Drain: An online log parsing approach
  with fixed depth tree,'' in \emph{ICWS'17: Proc. of the 24th International
  Conference on Web Services}, 2017.

\bibitem{he2018characterizing}
P.~He, Z.~Chen, S.~He, and M.~R. Lyu, ``Characterizing the natural language
  descriptions in software logging statements,'' in \emph{Proceedings of the
  33rd ACM/IEEE International Conference on Automated Software
  Engineering}.\hskip 1em plus 0.5em minus 0.4em\relax ACM, 2018, pp. 178--189.

\bibitem{He-DSN}
P.~He, J.~Zhu, S.~He, J.~Li, and M.~R. Lyu, ``An evaluation study on log
  parsing and its use in log mining,'' in \emph{46th Annual {IEEE/IFIP}
  International Conference on Dependable Systems and Networks, {DSN} 2016,
  Toulouse, France, June 28 - July 1, 2016}, 2016, pp. 654--661.

\bibitem{Drain}
P.~He, J.~Zhu, Z.~Zheng, and M.~R. Lyu, ``Drain: An online log parsing approach
  with fixed depth tree,'' in \emph{2017 {IEEE} International Conference on Web
  Services, {ICWS} 2017, Honolulu, HI, USA, June 25-30, 2017}, 2017, pp.
  33--40.

\bibitem{slheISSRE16}
S.~He, J.~Zhu, P.~He, and M.~Lyu, ``Experience report: System log analysis for
  anomaly detection,'' in \emph{ISSRE'16: Proc. of the 27th International
  Symposium on Software Reliability Engineering}, 2016.

\bibitem{log_survey}
S.~He, P.~He, Z.~Chen, T.~Yang, Y.~Su, and M.~R. Lyu, ``A survey on automated
  log analysis for reliability engineering,'' vol.~54, no.~6.\hskip 1em plus
  0.5em minus 0.4em\relax New York, NY, USA: Association for Computing
  Machinery, jul 2021.

\bibitem{he2018identifying}
S.~He, Q.~Lin, J.-G. Lou, H.~Zhang, M.~R. Lyu, and D.~Zhang, ``Identifying
  impactful service system problems via log analysis,'' in \emph{Proceedings of
  the 2018 26th ACM Joint Meeting on European Software Engineering Conference
  and Symposium on the Foundations of Software Engineering}, 2018, pp. 60--70.

\bibitem{log_empirical_study}
S.~He, X.~Zhang, P.~He, Y.~Xu, L.~Li, Y.~Kang, M.~Ma, Y.~Wei, Y.~Dang,
  S.~Rajmohan, and Q.~Lin, ``An empirical study of log analysis at microsoft,''
  in \emph{Proceedings of the 30th ACM Joint European Software Engineering
  Conference and Symposium on the Foundations of Software Engineering}, ser.
  ESEC/FSE 2022, 2022, p. 1465–1476.

\bibitem{huo2023evlog}
Y.~Huo, C.~Lee, Y.~Su, S.~Shan, J.~Liu, and M.~Lyu, ``Evlog: Evolving log
  analyzer for anomalous logs identification,'' \emph{arXiv preprint
  arXiv:2306.01509}, 2023.

\bibitem{huo2023semparser}
Y.~Huo, Y.~Su, C.~Lee, and M.~R. Lyu, ``Semparser: A semantic parser for log
  analytics,'' in \emph{2023 IEEE/ACM 45th International Conference on Software
  Engineering (ICSE)}.\hskip 1em plus 0.5em minus 0.4em\relax IEEE, 2023, pp.
  881--893.

\bibitem{jiang2008abstracting}
Z.~M. Jiang, A.~E. Hassan, P.~Flora, and G.~Hamann, ``Abstracting execution
  logs to execution events for enterprise applications (short paper),'' in
  \emph{2008 The Eighth International Conference on Quality Software}.\hskip
  1em plus 0.5em minus 0.4em\relax IEEE, 2008, pp. 181--186.

\bibitem{ll-31}
\BIBentryALTinterwordspacing
G.~Lee, J.~J. Lin, C.~Liu, A.~Lorek, and D.~V. Ryaboy, ``The unified logging
  infrastructure for data analytics at twitter,'' \emph{{PVLDB}}, vol.~5,
  no.~12, pp. 1771--1780, 2012. [Online]. Available:
  \url{http://vldb.org/pvldb/vol5/p1771\_georgelee\_vldb2012.pdf}
\BIBentrySTDinterwordspacing

\bibitem{li17ESEa}
H.~Li, T.~Chen, W.~Shang, and A.~E. Hassan, ``Studying software logging using
  topic models,'' \emph{Empirical Software Engineering}, 2017.

\bibitem{li17ESEb}
H.~Li, W.~Shang, and A.~E. Hassan, ``Which log level should developers choose
  for a new logging statement?'' \emph{Empirical Software Engineering},
  vol.~22, pp. 1684--1716, 2017.

\bibitem{liang2005filtering}
Y.~Liang, Y.~Zhang, A.~Sivasubramaniam, R.~K. Sahoo, J.~Moreira, and M.~Gupta,
  ``Filtering failure logs for a bluegene/l prototype,'' in \emph{2005
  International Conference on Dependable Systems and Networks (DSN'05)}.\hskip
  1em plus 0.5em minus 0.4em\relax IEEE, 2005, pp. 476--485.

\bibitem{liang2007failure}
Y.~Liang, Y.~Zhang, H.~Xiong, and R.~Sahoo, ``Failure prediction in ibm
  bluegene/l event logs,'' in \emph{the 7th IEEE International Conference on
  Data Mining (ICDM)}.\hskip 1em plus 0.5em minus 0.4em\relax IEEE, 2007, pp.
  583--588.

\bibitem{lim2014identifying}
M.-H. Lim, J.-G. Lou, H.~Zhang, Q.~Fu, A.~B.~J. Teoh, Q.~Lin, R.~Ding, and
  D.~Zhang, ``Identifying recurrent and unknown performance issues,'' in
  \emph{2014 IEEE International Conference on Data Mining (ICDM)}.\hskip 1em
  plus 0.5em minus 0.4em\relax IEEE, 2014, pp. 320--329.

\bibitem{MLC-11}
H.~Lin, J.~Zhou, B.~Yao, M.~Guo, and J.~Li, ``Cowic: {A} column-wise
  independent compression for log stream analysis,'' in \emph{15th {IEEE/ACM}
  International Symposium on Cluster, Cloud and Grid Computing (CCGrid)}, 2015,
  pp. 21--30.

\bibitem{lin2016log}
Q.~Lin, H.~Zhang, J.-G. Lou, Y.~Zhang, and X.~Chen, ``Log clustering based
  problem identification for online service systems,'' in \emph{Proceedings of
  the 38th International Conference on Software Engineering Companion}.\hskip
  1em plus 0.5em minus 0.4em\relax ACM, 2016, pp. 102--111.

\bibitem{liu2008isolation}
F.~T. Liu, K.~M. Ting, and Z.-H. Zhou, ``Isolation forest,'' in \emph{the 8th
  IEEE International Conference on Data Mining (ICDM)}.\hskip 1em plus 0.5em
  minus 0.4em\relax IEEE, 2008, pp. 413--422.

\bibitem{liu2023incident}
J.~Liu, S.~He, Z.~Chen, L.~Li, Y.~Kang, X.~Zhang, P.~He, H.~Zhang, Q.~Lin,
  Z.~Xu \emph{et~al.}, ``Incident-aware duplicate ticket aggregation for cloud
  systems,'' \emph{arXiv preprint arXiv:2302.09520}, 2023.

\bibitem{liu2023scalable}
J.~Liu, J.~Huang, Y.~Huo, Z.~Jiang, J.~Gu, Z.~Chen, C.~Feng, M.~Yan, and M.~R.
  Lyu, ``Scalable and adaptive log-based anomaly detection with expert in the
  loop,'' \emph{arXiv preprint arXiv:2306.05032}, 2023.

\bibitem{liu2019logzip}
J.~Liu, J.~Zhu, S.~He, P.~He, Z.~Zheng, and M.~R. Lyu, ``Logzip: Extracting
  hidden structures via iterative clustering for log compression,'' in
  \emph{2019 34th IEEE/ACM International Conference on Automated Software
  Engineering (ASE)}.\hskip 1em plus 0.5em minus 0.4em\relax IEEE, 2019, pp.
  863--873.

\bibitem{uniparser}
Y.~Liu, X.~Zhang, S.~He, H.~Zhang, L.~Li, Y.~Kang, Y.~Xu, M.~Ma, Q.~Lin,
  Y.~Dang, S.~Rajmohan, and D.~Zhang, ``Uniparser: A unified log parser for
  heterogeneous log data,'' in \emph{Proceedings of the ACM Web Conference
  2022}, ser. WWW '22.\hskip 1em plus 0.5em minus 0.4em\relax Association for
  Computing Machinery, 2022, p. 1893–1901.

\bibitem{lou2010mining}
J.-G. Lou, Q.~Fu, S.~Yang, Y.~Xu, and J.~Li, ``Mining invariants from console
  logs for system problem detection.'' in \emph{USENIX Annual Technical
  Conference}, 2010, pp. 1--14.

\bibitem{IPLoM}
A.~Makanju, A.~N. Zincir{-}Heywood, and E.~E. Milios, ``Clustering event logs
  using iterative partitioning,'' in \emph{Proceedings of the 15th {ACM}
  {SIGKDD} International Conference on Knowledge Discovery and Data Mining,
  Paris, France, June 28 - July 1, 2009}, 2009, pp. 1255--1264.

\bibitem{IRbook08}
C.~Manning, P.~Raghavan, and H.~Schutze, \emph{Introduction to Information
  Retrieval}.\hskip 1em plus 0.5em minus 0.4em\relax Cambridge University
  Press, 2008.

\bibitem{messaoudi2018search}
S.~Messaoudi, A.~Panichella, D.~Bianculli, L.~Briand, and R.~Sasnauskas, ``A
  search-based approach for accurate identification of log message formats,''
  in \emph{Proceedings of the 26th Conference on Program Comprehension}.\hskip
  1em plus 0.5em minus 0.4em\relax ACM, 2018, pp. 167--177.

\bibitem{mi2013toward}
H.~Mi, H.~Wang, Y.~Zhou, M.~R.-T. Lyu, and H.~Cai, ``Toward fine-grained,
  unsupervised, scalable performance diagnosis for production cloud computing
  systems,'' \emph{IEEE Transactions on Parallel and Distributed Systems},
  vol.~24, no.~6, pp. 1245--1255, 2013.

\bibitem{SHISO}
M.~Mizutani, ``Incremental mining of system log format,'' in \emph{2013 {IEEE}
  International Conference on Services Computing, Santa Clara, CA, USA, June 28
  - July 3, 2013}, 2013, pp. 595--602.

\bibitem{anomaly2}
\BIBentryALTinterwordspacing
P.~Mosendz. (2014) When it goes down, facebook loses \$24,420 per minute.
  [Online]. Available:
  \url{https://www.theatlantic.com/technology/archive/2014/10/facebook-is-losing-24420-per-minute/382054/}
\BIBentrySTDinterwordspacing

\bibitem{nagappan2010abstracting}
M.~Nagappan and M.~A. Vouk, ``Abstracting log lines to log event types for
  mining software system logs,'' in \emph{2010 7th IEEE Working Conference on
  Mining Software Repositories (MSR)}, 2010, pp. 114--117.

\bibitem{BGLdata}
A.~Oliner and J.~Stearley, ``What supercomputers say: A study of five system
  logs,'' in \emph{DSN}, 2007.

\bibitem{pecchia2015industry}
A.~Pecchia, M.~Cinque, G.~Carrozza, and D.~Cotroneo, ``Industry practices and
  event logging: Assessment of a critical software development process,'' in
  \emph{Proceedings of the 37th International Conference on Software
  Engineering-Volume 2}.\hskip 1em plus 0.5em minus 0.4em\relax IEEE Press,
  2015, pp. 169--178.

\bibitem{MLC-18}
B.~R{\'{a}}cz and A.~Luk{\'{a}}cs, ``High density compression of log files,''
  in \emph{2004 Data Compression Conference {(DCC} 2004), 23-25 March 2004,
  Snowbird, UT, {USA}}, 2004, p. 557.

\bibitem{rakha2018revisiting}
M.~S. Rakha, C.-P. Bezemer, and A.~E. Hassan, ``Revisiting the performance
  evaluation of automated approaches for the retrieval of duplicate issue
  reports,'' \emph{IEEE Transactions on Software Engineering}, vol.~44, no.~12,
  pp. 1245--1268, 2018.

\bibitem{rodrigues2021clp}
K.~Rodrigues, Y.~Luo, and D.~Yuan, ``$\{$CLP$\}$: Efficient and scalable search
  on compressed text logs,'' in \emph{15th $\{$USENIX$\}$ Symposium on
  Operating Systems Design and Implementation ($\{$OSDI$\}$ 21)}, 2021, pp.
  183--198.

\bibitem{scholkopf2001estimating}
B.~Sch{\"o}lkopf, J.~C. Platt, J.~Shawe-Taylor, A.~J. Smola, and R.~C.
  Williamson, ``Estimating the support of a high-dimensional distribution,''
  \emph{Neural Computation}, vol.~13, no.~7, pp. 1443--1471, 2001.

\bibitem{verification}
W.~Shang, Z.~M. Jiang, H.~Hemmati, B.~Adams, A.~E. Hassan, and P.~Martin,
  ``Assisting developers of big data analytics applications when deploying on
  hadoop clouds,'' in \emph{35th International Conference on Software
  Engineering, {ICSE} '13, San Francisco, CA, USA, May 18-26, 2013}, 2013, pp.
  402--411.

\bibitem{shima2016length}
K.~Shima, ``Length matters: Clustering system log messages using length of
  words,'' \emph{arXiv preprint arXiv:1611.03213}, 2016.

\bibitem{LogSig}
L.~Tang, T.~Li, and C.~Perng, ``Logsig: generating system events from raw
  textual logs,'' in \emph{Proceedings of the 20th {ACM} Conference on
  Information and Knowledge Management, {CIKM} 2011, Glasgow, United Kingdom,
  October 24-28, 2011}, 2011, pp. 785--794.

\bibitem{anomaly1}
\BIBentryALTinterwordspacing
UpGuard. (2016) The cost of downtime at the world's biggest online retailer.
  [Online]. Available:
  \url{https://www.upguard.com/blog/the-cost-of-downtime-at-the-worlds-biggest-online-retailer}
\BIBentrySTDinterwordspacing

\bibitem{SLCT}
R.~Vaarandi, ``A data clustering algorithm for mining patterns from event
  logs,'' in \emph{IP Operations \& Management, 2003.(IPOM 2003). 3rd IEEE
  Workshop on}.\hskip 1em plus 0.5em minus 0.4em\relax IEEE, 2003, pp.
  119--126.

\bibitem{vaarandi2015logcluster}
R.~Vaarandi and M.~Pihelgas, ``Logcluster-a data clustering and pattern mining
  algorithm for event logs,'' in \emph{2015 11th International Conference on
  Network and Service Management (CNSM)}.\hskip 1em plus 0.5em minus
  0.4em\relax IEEE, 2015, pp. 1--7.

\bibitem{wang2022spine}
X.~Wang, X.~Zhang, L.~Li, S.~He, H.~Zhang, Y.~Liu, L.~Zheng, Y.~Kang, Q.~Lin,
  Y.~Dang \emph{et~al.}, ``Spine: a scalable log parser with feedback
  guidance,'' in \emph{Proceedings of the 30th ACM Joint European Software
  Engineering Conference and Symposium on the Foundations of Software
  Engineering}, 2022, pp. 1198--1208.

\bibitem{spine}
X.~Wang, X.~Zhang, L.~Li, S.~He, H.~Zhang, Y.~Liu, L.~Zheng, Y.~Kang, Q.~Lin,
  Y.~Dang, S.~Rajmohan, and D.~Zhang, ``Spine: A scalable log parser with
  feedback guidance,'' in \emph{Proceedings of the 30th ACM Joint European
  Software Engineering Conference and Symposium on the Foundations of Software
  Engineering}, ser. ESEC/FSE 2022.\hskip 1em plus 0.5em minus 0.4em\relax
  Association for Computing Machinery, 2022, p. 1198–1208.

\bibitem{wei2023loggrep}
J.~Wei, G.~Zhang, J.~Chen, Y.~Wang, W.~Zheng, T.~Sun, J.~Wu, and J.~Jiang,
  ``Loggrep: Fast and cheap cloud log storage by exploiting both static and
  runtime patterns,'' in \emph{Proceedings of the Eighteenth European
  Conference on Computer Systems}, 2023, pp. 452--468.

\bibitem{wei2021feasibility}
J.~Wei, G.~Zhang, Y.~Wang, Z.~Liu, Z.~Zhu, J.~Chen, T.~Sun, and Q.~Zhou, ``On
  the feasibility of parser-based log compression in $\{$Large-Scale$\}$ cloud
  systems,'' in \emph{19th USENIX Conference on File and Storage Technologies
  (FAST 21)}, 2021, pp. 249--262.

\bibitem{xu2023hue}
J.~Xu, Q.~Fu, Z.~Zhu, Y.~Cheng, Z.~Li, Y.~Ma, and P.~He, ``Hue: A user-adaptive
  parser for hybrid logs,'' \emph{arXiv preprint arXiv:2308.07085}, 2023.

\bibitem{Xu_sosp_2009}
W.~Xu, L.~Huang, A.~Fox, D.~A. Patterson, and M.~I. Jordan, ``Detecting
  large-scale system problems by mining console logs,'' in \emph{SOSP}, 2009,
  pp. 117--132.

\bibitem{ll-42}
------, ``Detecting large-scale system problems by mining console logs,'' in
  \emph{Proceedings of the 27th International Conference on Machine Learning
  (ICML-10), June 21-24, 2010, Haifa, Israel}, 2010, pp. 37--46.

\bibitem{yu2023logreducer}
G.~Yu, P.~Chen, P.~Li, T.~Weng, H.~Zheng, Y.~Deng, and Z.~Zheng, ``Logreducer:
  Identify and reduce log hotspots in kernel on the fly,'' in \emph{2023
  IEEE/ACM 45th International Conference on Software Engineering (ICSE)}.\hskip
  1em plus 0.5em minus 0.4em\relax IEEE, 2023, pp. 1763--1775.

\bibitem{yuan2012conservative}
D.~Yuan, S.~Park, P.~Huang, Y.~Liu, M.~M. Lee, X.~Tang, Y.~Zhou, and S.~Savage,
  ``Be conservative: enhancing failure diagnosis with proactive logging,'' in
  \emph{Presented as part of the 10th $\{$USENIX$\}$ Symposium on Operating
  Systems Design and Implementation ($\{$OSDI$\}$ 12)}, 2012, pp. 293--306.

\bibitem{yuan2012characterizing}
D.~Yuan, S.~Park, and Y.~Zhou, ``Characterizing logging practices in
  open-source software,'' in \emph{Proceedings of the 34th International
  Conference on Software Engineering}.\hskip 1em plus 0.5em minus 0.4em\relax
  IEEE Press, 2012, pp. 102--112.

\bibitem{onion}
X.~Zhang, Y.~Xu, S.~Qin, S.~He, B.~Qiao, Z.~Li, H.~Zhang, X.~Li, Y.~Dang,
  Q.~Lin, M.~Chintalapati, S.~Rajmohan, and D.~Zhang, ``Onion: Identifying
  incident-indicating logs for cloud systems,'' in \emph{Proceedings of the
  29th ACM Joint Meeting on European Software Engineering Conference and
  Symposium on the Foundations of Software Engineering}, ser. ESEC/FSE
  2021.\hskip 1em plus 0.5em minus 0.4em\relax Association for Computing
  Machinery, 2021, p. 1253–1263.

\bibitem{MTracer}
J.~Zhou, Z.~Chen, H.~Mi, and J.~Wang, ``Mtracer: {A} trace-oriented monitoring
  framework for medium-scale distributed systems,'' in \emph{8th {IEEE}
  International Symposium on Service Oriented System Engineering ({SOSE})},
  2014, pp. 266--271.

\bibitem{TraceBench}
J.~Zhou, Z.~Chen, J.~Wang, Z.~Zheng, and M.~R. Lyu, ``Trace bench: An open data
  set for trace-oriented monitoring,'' in \emph{{IEEE} 6th International
  Conference on Cloud Computing Technology and Science (CloudCom)}, 2014, pp.
  519--526.

\bibitem{zhu15learningToLog}
J.~Zhu, P.~He, Q.~Fu, H.~Zhang, M.~R. Lyu, and D.~Zhang, ``Learning to log:
  Helping developers make informed logging decisions,'' in \emph{ICSE}, 2015.

\bibitem{zhu2018tools}
J.~Zhu, S.~He, J.~Liu, P.~He, Q.~Xie, Z.~Zheng, and M.~R. Lyu, ``Tools and
  benchmarks for automated log parsing,'' in \emph{Proceedings of the 41st
  International Conference on Software Engineering ({ICSE-SEIP})}, 2019, pp.
  121--130.

\end{thebibliography}
